\documentclass{aa}

\usepackage{graphicx,float}
\usepackage{psfig}
\usepackage{epsfig}

\def \sax {BeppoSAX}

\def \hcm {\hbox {\ifmmode $ atom cm$^{-2}\else atom cm$^{-2}$\fi}}
\def \arcmin {\hbox{$^\prime$}}

\def\approxgt{\mathrel{\hbox{\rlap{\lower.55ex \hbox {$\sim$}}
        \kern-.3em \raise.4ex \hbox{$>$}}}}
\def\approxlt{\mathrel{\hbox{\rlap{\lower.55ex \hbox {$\sim$}}
        \kern-.3em \raise.4ex \hbox{$<$}}}}

\begin{document}

%\thesaurus{   }  

\title{BeppoSAX survey of Be/X-ray binary candidates}

\author{J. M. Torrej\'{o}n\inst{1}
        \and A. Orr\inst{2} 
}
\offprints{J. M. Torrej\'{o}n (jmt@disc.ua.es)}

\institute{ University of Alicante, Department of Physics, EPS, Ap. 99, 
            E-03080, Alicante, Spain
\and
       Astrophysics Division, Space Science Department of ESA, ESTEC,
       Postbus 299, NL-2200 AG Noordwijk, The Netherlands
}
\date{Received ; Accepted: 200? Month dy}

%\markboth{BeppoSAX survey of Be/X-ray binaries}{BeppoSAX 
% survey of Be/X-ray binaries}

\abstract{We present a BeppoSAX survey of five Be/X-ray binary
  candidates. We report on the identification of two of them, HD~110432 and
  HD~141926, as low luminosity Be/X-ray binaries. For HD~110432 we report
  on the detection of a pulsation period of $\sim $14 ks. Because the
  luminosity of these sources is low and their spectra do not require
  non-thermal emission models, these systems are good Be+White Dwarf
  candidates. If the pulsation period for HD 110432 is confirmed, this
  system would be the most firm Be+WD  candidate found up to date. The
  other three objects HD~65663, HD~249179 and BD+53~2262 did not show
  detectable X-ray emission. We argue that, while the properties of
  BD+53~2262 are still consistent with a quiescent Be+Neutron Star
  scenario, the lack of detection for the other two objects implies that
  they are most probably not X-ray binaries.   
} 

\maketitle

\section{Introduction}
\label{sect:intro}
During the {\it HEAO-1} all-sky survey with the Modulation Collimator, Tuohy et
al. (1988) observed several faint unidentified hard X-ray sources. These
sources were associated with emission line stars located within the
positional error boxes of the X-ray experiment, and consequently proposed
as Be/X-ray binary candidates (BeXRBs). However, as warned by these
authors, the relatively large error boxes could make some of these
identifications uncertain. These sources are included in the catalogues of
High Mass X-ray binaries (Liu et al. 2000). However, no observation
  has been
reported since then with other X-ray telescopes and, consequently, their very nature as X-ray sources has not been firmly established.

BeXRBs are a major subclass of High Mass X-ray binaries (HMXRBs) where a 
compact object orbits a Be star in an eccentric orbit (see Apparao 1994,
Coe 2000 for recent reviews). A Be star is a star of spectral type B and
luminosity class III to V, whose spectrum shows Balmer lines in emission as
well as a strong infrared excess (Slettebak 1988). These two characteristic
features seem to arise from a circumstellar envelope, whose very origin
remains unclear. This envelope consists of a high density low velocity
component confined at the equatorial regions and a low density high
velocity wind produced mostly on the polar regions of the star (Lamers and
Waters, 1987). The compact object interacts with this envelope and accretes
matter thereby producing X-rays. Most of the soft X-rays are probably
absorbed in the envelope itself (Apparao 1994) and therefore the observed
X-ray spectrum is usually hard, the most conspicuous emission being within
the range of 2--20~keV.

The compact object is generally thought to be a neutron star (Be+NS). The
X-ray luminosities usually found in BeXRBs ($\sim 10^{36-37}$ erg
s$^{-1}$) and the limited range in spectral types covered by the optical
counterparts of these objects (Van den Heuvel and Rappaport 1987,
Negueruela 1998) confirm this picture. Evolutionary calculations show,
however, that there should be a large number of less massive binary systems
with white dwarf accretors (Waters et al. 1989, Raguzova 2001). These
systems would present far ultraviolet emission and could be observable as
low luminosity X-ray emitters ($L_{\rm X}\sim 10^{32}$ erg s$^{-1}$). To date however, only three B+White Dwarf  systems have been found and no Be+White Dwarf (Be+WD) system has been firmly established (see however Kurakami et al. 1986, Haberl 1995, and also Robinson \& Smith 2000, for the case of $\gamma$~Cas). 

In order to assess the nature of these BeXRBs we have performed an analysis of unpublished archive data from the BeppoSAX Narrow Field Instruments for five of Tuohy's BeXRB candidates.

\begin{table*}
\caption[]{Our sample. T$_{\rm int}$ is the net MECS integration time in ks.
The spectral types have been taken from Liu et al. (2000). NA: not available}
\begin{flushleft}
\begin{tabular}{lrrlrrc}
\hline\noalign{\smallskip}
Source    & R.A.$_{2000}$ & Dec.$_{2000}$ & Sp. type & Obs. start & Obs. end &  T$_{\rm int}$ (ks) \\

\noalign{\smallskip\hrule\smallskip}

HD 249179 & 05 55 55.05 & +28 47 06.4 & B5ne & 19961004 05:02 &  19961004 12:10 & 20.9\\
HD 65663  & 07 56 15.77 & -61 05 58.0  & B8IIIe & 19961204 09:55 &  19961204 15:36 & 47.0 \\
HD 110432 & 12 42 50.27 & -63 03 31.0 & B0.5IIIe & 19970114 11:27 &  19970114 17:14 & 16.0 \\
HD 141926 & 15 54 21.82 & -55 19 44.8 & B2nne & 19980202 03:44 &  19980202 10:32 & 18.2 \\
BD+53 2262 & 19 32 52.31 & +53 52 45.5 & NA  & 19970504 15:21 &  19970504 21:36 & 17.8 \\
\noalign{\smallskip\hrule\smallskip}
\end{tabular}
\end{flushleft}
\label{tab:obs_times}
\end{table*}

\section{Observations}
\label{sect:obs}

Results from the Low-Energy Concentrator Spectrometer (LECS;
0.1--10~keV; Parmar et al. \cite{p:97}), the Medium-Energy Concentrator
Spectrometer (MECS; 1.8--10~keV; Boella et al. \cite{b:97})
 and the Phoswich Detection System (PDS; 15--300~keV; Frontera et al. 
\cite{f:97}) on-board \sax\
are presented. All these instruments are coaligned and  referred
to as Narrow Field Instruments, or NFI.
The MECS consists of two (--three until 1997 May 9) grazing incidence
telescopes with imaging gas scintillation proportional counters in
their focal planes. The LECS uses an identical concentrator system as
the MECS, but utilizes an ultra-thin entrance window and
a driftless configuration to extend the low-energy response to
0.1~keV.

The non-imaging PDS consists of four independent units arranged in 
pairs each having a separate collimator. Each collimator was alternatively
rocked on-source and 210\arcmin\ off-source every 96~s during the observation.

The data from the High Pressure Gas Scintillation Proportional Counter
on-board \sax\
(HPGSPC; 5--120~keV; Manzo et al. \cite{m:97})
were not used in the present study because none of the sources
were bright enough for this instrument.

Table \ref{tab:obs_times} lists the \sax\  observation epochs and net 
exposure times for our sample.

Good data were selected from intervals when the elevation angle
above the Earth's limb was $>$$4^{\circ}$ and when the instrument
configurations were nominal, using the SAXDAS 2.0.0 data analysis package.

In order to produce the spectra, LECS and MECS data were extracted 
centered on the position of each source using radii of 4\arcmin\ 
and 2\arcmin, respectively. These are smaller than the standard recommended
radii, but they are justified in the case of faint sources.
However, the light curve of HD~110432 (Fig. 1) was produced using a ``full''
4\arcmin\ extraction radius in order to exclude any effects due to 
possible spacecraft ``wobbling''. 

The PDS data are only useful in the case of our brightest source 
HD~110432. PDS event filtering was made using the so-called PDS 
``Variable Rise Time'' selection method. This method 
increases the signal to noise ratio of the PDS data in weak 
sources such as ours.
Background subtraction for the PDS was performed in the standard way using data
obtained during intervals when the collimators were offset from the 
source. 

Background subtraction for the imaging instruments (LECS and MECS)
was performed using the appropriate standard files.

\section{Results}

The 2--10 keV fluxes and flux upper limits for our sample are listed in 
Table \ref{tab:fluxes}.  All measured results are quoted with 90\% confidence uncertainty 
intervals, for one parameter of interest, unless otherwise quoted.

The uncertainty on the 2--10 keV fluxes is a combination of the statistical 
uncertainty on the fit (given in Table 2) and systematic effects. 
In the case of our relatively faint sources the systematic component 
is dominated by the background subtraction.
By comparing different MECS and LECS background subtraction techniques,
involving local and standard backgrounds, and by testing different source
extraction radii, it appears that the largest systematic flux deviations
are no larger than 2\% of the measured fluxes.

\begin{table}
\caption[]{2-10 keV fluxes measured with \sax, in units of $10^{-12}$ erg 
s$^{-1}$ cm$^{-2}$. The fluxes are quoted with their statistical
  errors at 90\% confidence level.}
\begin{flushleft}
\begin{tabular}{lc}
\hline\noalign{\smallskip}
Source & F$_{2-10}$ \\
\noalign{\smallskip\hrule\smallskip}

HD 249179  & $<$ 0.19\\
HD 65663   & $<$ 0.14\\
HD 110432  & $22.70\pm 0.01$ \\
HD 141926  & $1.12\pm 0.04$ \\
BD+53 2262 & $<$ 0.16\\
\noalign{\smallskip\hrule\smallskip}
\end{tabular}
\end{flushleft}
\label{tab:fluxes}
\end{table}

\subsection{HD~110432 - lightcurve and spectrum}
\label{subsect:HD11}

\begin{figure}
 \centerline{\psfig{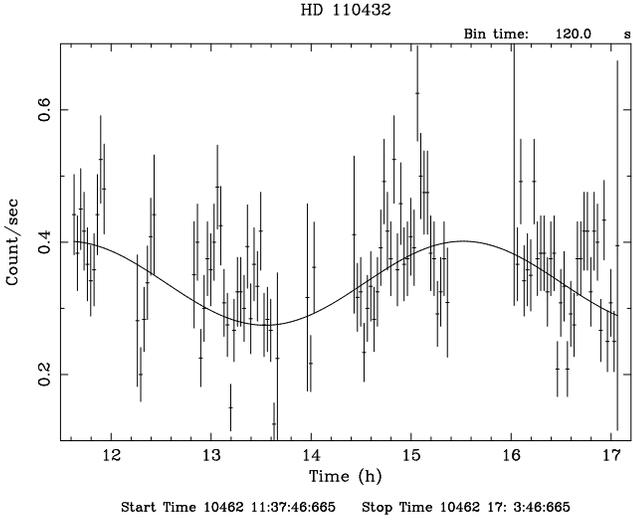}}
  \caption[]{1.8--10 keV MECS background subtracted lightcurve of
HD~110432. The data are binned in 120 s intervals. }
  \label{fig:lightcurve}
\end{figure}

The MECS 1.8--10 keV background-substracted barycentric corrected
light curve of HD~110432, is shown in Fig. \ref{fig:lightcurve}.
The background subtraction was made using the standard composite
MECS background event file.   
Although the resulting curve is relatively
noisy, a constant X-ray flux can definitely be rejected with a confidence
level greater than 99\%, the fit yielding a $\chi^2 = 277$ (dof = 105,
dof=degrees of freedom). 
A linear function can equally be excluded ($\chi^2 = 277$, dof = 104). A better fit ($\chi^2 = 214$, dof=102) is obtained with a sinusoidal curve of the type:
\begin{equation}
A+B\sin(2\pi(\tau-\phi)/P)
\end{equation}

where A and B are constants, $\tau$ is the time, $\phi$ is the phase and P the period. The best fit parameters are presented in table \ref{tab:par}.

\begin{table}
\caption[]{Light curve parameters for HD~110432 measured with \sax ~using a time binning of 120 s.}
\begin{flushleft}
\begin{tabular}{ccc}
\hline\noalign{\smallskip}
Parameter & Value & \\
\noalign{\smallskip\hrule\smallskip}

A  & 0.34$\pm$ 0.01 & c~s$^{-1}$\\
B  & 0.06 $\pm$ 0.02  & c~s$^{-1}$\\
$\phi$ & 5589$\pm$ 820 & s  \\
P& (1.42 $\pm 0.14)\times 10^{4}$ & s\\

\noalign{\smallskip\hrule\smallskip}
\end{tabular}
\end{flushleft}
\label{tab:par}
\end{table}

\begin{figure}
 \centerline{\psfig{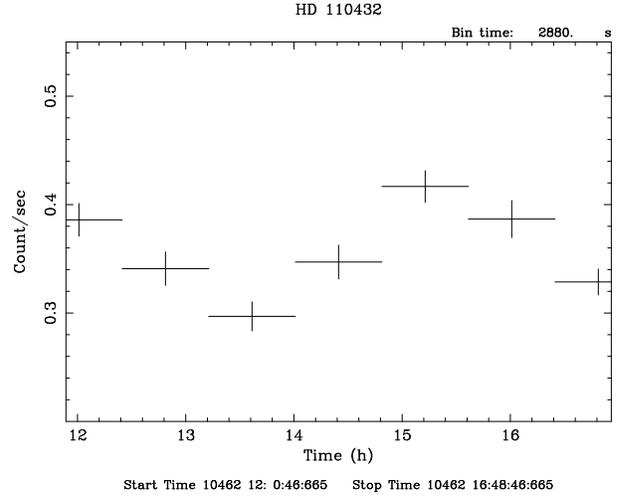}}
  \caption[]{The same as in Fig.1 but with a time binning of 2880 s. Statistically significant variations from bin to bin are clearly seen.}
  \label{fig:lightcurve2}
\end{figure}

Several binning times where tried to perform the fitting. The parameters, in particular the period P, remained essentially unchanged. Therefore, the sinusoidal pulsation is not an artifact of the binning process. A binning time of 120 s was chosen in order to get a sufficient number of counts per bin to apply $\chi^2$ statistics while maintaining good timing resolution. Significant short time scale variations (of the order of minutes) can be seen. To see more clearly the underlying sinusoidal modulation we have tried a binning time of 2880 s which would sample the source twice per satellite orbit. The resulting light curve is plotted in Fig. \ref{fig:lightcurve2}. Statistically significant variations are clearly seen from bin to bin.

In order to further reject the possibility of the modulation being due to satellite wobbling a light curve analysis from a background ring around the source was performed. The result is completely compatible ($\chi^2 = 113$, dof=105) with a constant value ($0.26\pm 0.01$~c~s$^{-1}$). Therefore we conclude that the period must be associated with intrinsic variability of the source. The deduced period ($\sim 14~\rm ks$) is only slightly shorter than the observing window ($\sim 19~\rm ks$) so that no clear phase overlap could be established. A longer observation is clearly needed to confirm it.

The overall spectrum of HD~110432  was investigated by simultaneously
fitting data from  the BeppoSAX NFI, using XSPEC (Arnaud, 1996).
The LECS and MECS spectra were rebinned to oversample the full
width half maximum of the energy resolution by
a factor 3 and to have additionally a minimum of 20 counts 
per bin to allow use of the $\chi^2$ statistic. 
The  PDS 
spectra were rebinned using the standard techniques in SAXDAS.
Data were selected in the energy ranges
0.1--10.0~keV (LECS), 1.65--10~keV (MECS)
and 15--70~keV (PDS) 
where the instrument responses are well determined and sufficient
counts obtained. 
This gives
background-subtracted count rates of 0.064, 0.363,  and 0.147 ~s$^{-1}$ 
for the LECS, MECS  and PDS, respectively.

The photoelectric absorption
cross sections of Morrison \& McCammon (\cite{m:83}) and the
solar abundances of Anders \& Grevesse (\cite{a:89}) are used throughout.

Factors were included in the spectral fitting to allow for normalisation 
differences between the instruments. These factors were constrained
to be within their recommended ranges during the fitting. 

\begin{figure}
 \centerline{\psfig{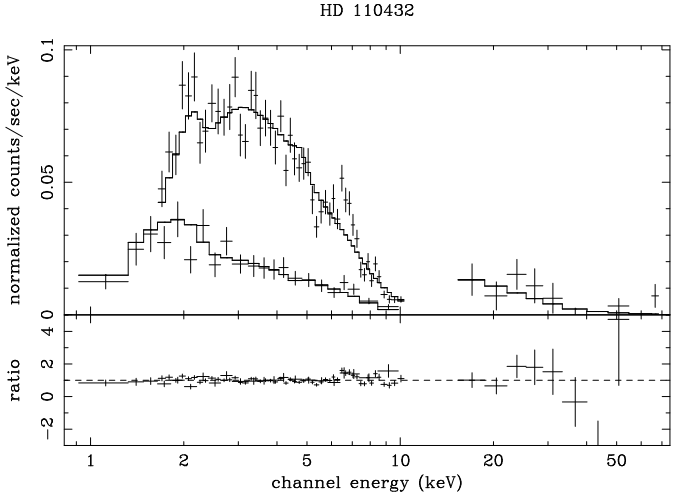}}
  \caption[]{LECS, MECS and PDS 0.9--50 keV count rate 
  spectrum of HD~110432, with data-to-model ratio in lower plot. 
  The fitted model is a cutoff power-law with neutral photo-electric
absorption. Positive residuals are clearly seen between 6--7 keV, and
 7--8 keV.  
 }
  \label{fig:hd11-spectrum}
\end{figure}

Initially, simple models were tried, including absorbed power-law,
 and cutoff power-law models. A power-law with a photon index $\alpha = 1.77\pm 0.08$ 
gives an unacceptable fit with $\chi ^2$ of 178.77 for 95 dof. A cutoff power-law model with $\alpha = 1.35\pm0.01$ and ${\rm E_{cutoff} = 16.69}$~keV
gives a better fit with a $\chi ^2$ of
164.17 for 94 dof. Fig. \ref{fig:hd11-spectrum} shows the \sax\ spectrum of HD~110432
and the residuals to the cutoff power-law fit. 
Positive residuals are clearly seen around 6--7 keV and 8--9 keV, and
a possible broad emission feature between  20--30 keV. The addition of Gaussian lines at 6.76 and 8.41 keV significantly improves the fit, giving a $\chi ^2$ of 114.89 for 91 dof for the addition of one line and     
$\chi ^2$ of 107.13 for 89 dof with the addition of two lines, respectively.

A comparably good fit, however,  is obtained with a MEKAL interpretation. The spectrum can be well fitted with an optically thin plasma at $kT=10.55$ ~keV ($\chi ^2$ of 114.93 for 93 dof). The parameters for the different models are shown in table \ref{tab:mod} (only models with $\chi ^2/\nu < 2$ are included).

The observed 2--10 keV flux is F$_{2-10}$ is 22.70$\times 10^{-12}$
erg cm$^{-2}$ s$^{-1}$ for the cutoff power-law + iron line model and
22.88$\times 10^{-12}$ erg cm$^{-2}$ s$^{-1}$ for the MEKAL model, respectively.

\begin{table}
\caption[]{Model parameters for HD~110432}
\begin{flushleft}
\begin{tabular}{lr}
\hline\noalign{\smallskip}
Parameter & Value \\
\noalign{\smallskip\hrule\smallskip}

{\bf cutoffpl} &  \\
 & \\
 N$_{\rm H}(10^{22} \rm cm^{-2})$& 1.05 $\pm$ 0.01\\
$\alpha$ & 1.35$\pm$ 0.01   \\
E$_{\rm cutoff}$~(keV) & 16.7$\pm^{35.6}_{9.5}$ \\
$\chi^{2}_{\rm r}$(dof) & 1.75(94)\\
 & \\
{\bf  cutoffpl + one line }& \\
 & \\
N$_{\rm H}(10^{22} \rm cm^{-2})$& 1.25 $\pm $0.30\\
$\alpha$ & 1.56$\pm$ 0.27   \\
E$_{\rm cutoff}$~(keV) & 21.07$\pm^{60.0}_{15.5} $ \\
E$_{\rm line}$~(keV) & 6.77$\pm$ 0.10     \\
$\sigma_{\rm line}$ & 0.23$\pm$ 0.15 \\
EW ~(eV)& 598  $\pm$ 340\\
$\chi^{2}_{\rm r}$(dof) & 1.26(91)\\
 & \\
{\bf cutoffpl + two lines }& \\
 & \\
N$_{\rm H}(10^{22} \rm cm^{-2})$& 1.38 $\pm $0.30\\
$\alpha$ & 1.63$\pm$ 0.27   \\
E$_{\rm cutoff}$~(keV) & 19.9$\pm^{17.0}_{9.5}$ \\
E$_{\rm line 1}$~(keV) & 6.76$\pm$ 0.10     \\
$\sigma_{\rm line1}$ & 0.26$\pm$ 0.15 \\
EW$_{1}$ ~(eV) & 716 $\pm 430$\\
E$_{\rm line 2}$~(keV) & 8.41$\pm$ 0.10     \\
$\sigma_{\rm line2}$ & 0.10$\pm$ 0.05 \\
EW$_{2}$ ~(eV)& 340 $\pm$ 204 \\
$\chi^{2}_{\rm r}$(dof) & 1.20(89)\\
 & \\
{\bf MEKAL} & \\
 & \\
N$_{\rm H}(10^{22} \rm cm^{-2})$& 1.08 $\pm$ 0.30\\
kT~(keV) & 10.55$\pm 1.90$  \\
Abundance & 0.78$\pm 0.20$ \\
$\chi^{2}_{\rm r}$(dof) & 1.24(93)\\

\noalign{\smallskip\hrule\smallskip}
\end{tabular}
\end{flushleft}
\label{tab:mod}
\end{table}

\begin{figure}
 \centerline{\psfig{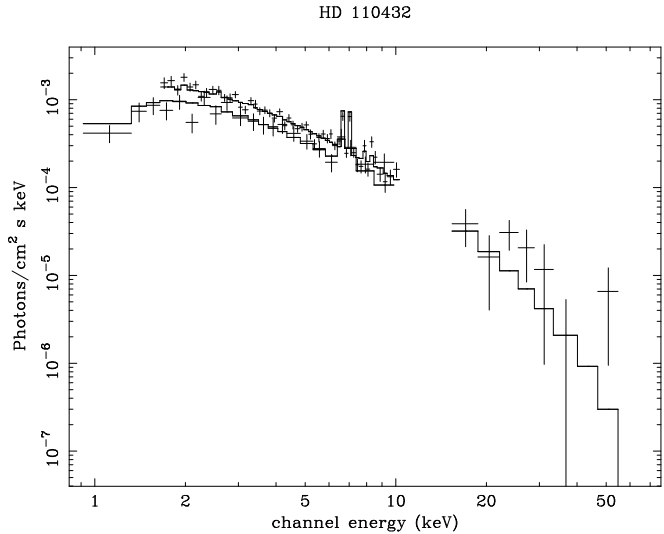}}
  \caption[]{The unfolded spectrum of HD 110432 with the MEKAL fitted model. Note the presence of the emission lines between 6--7 keV, and 8--9 keV.  
 }
  \label{fig:hd11b-spectrum}
\end{figure}

\subsection{HD~141926 -  spectrum}
\label{subsect:HD14:sp}

Only LECS and MECS data were used for this relatively faint source. 
The data were selected in the energy ranges 1.4-9.6 keV and 1.5-12 keV, respectively,  giving 
background-subtracted count rates of 0.004  and 0.010  ~s$^{-1}$ 
for the LECS and MECS, respectively. 

\begin{figure}
 \centerline{\psfig{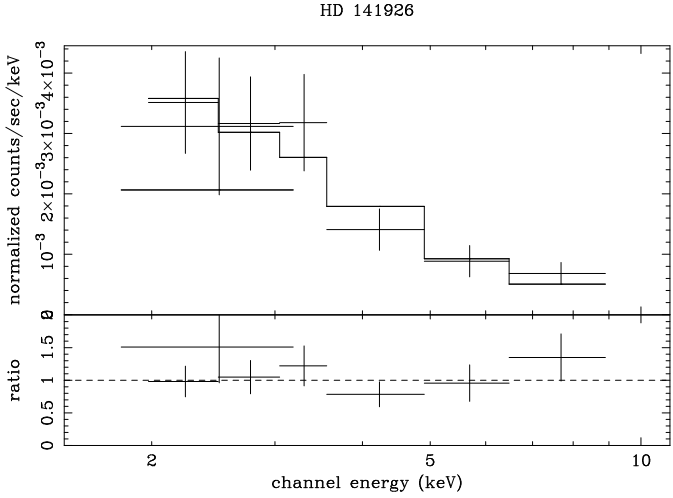}}
  \caption[]{LECS and MECS 2--10 keV count rate 
  spectrum of HD~141926, with data-to-model ratio in lower plot. The fitted model is a Raymond emission of hot ionized plasma at kT=4.08 keV with neutral photo-electric absorption}
  \label{fig:hd14-spectrum}
\end{figure}

In order to investigate its X-ray spectrum, we initially tried simple models. Their results are listed in Table 5. The absorbed cutoff power-law gives a very poor fit ($\chi^2 = 5.91$ for 3 dof). A simple power law gives a slightly better but still poor fit ($\chi^2 = 5.80$ for 4 dof) with photon index $\alpha=2.12 \pm 0.02$ and an undefined absorption column with an upper limit of 2.73$\times 10^{22}$ atoms cm$^{-2}$. The X-ray spectrum of HD~141926 can, however, be well described by an emission spectrum of hot, diffuse gas with a temperature of $\sim 4.5$ ~keV either with MEKAL interpretation ($kT=4.53$ ~keV, $\chi^{2}=3.49$ for 4 dof) or a Raymond model interpretation ($kT=4.47$ ~keV, $\chi^{2}=3.28$ for 4 dof). The abundances in both cases have been frozen to solar. Both models yield, however, unphysically low values for the absorption column. Using additional optical data to pin down the amount of absorption, we have been able to fit rather satisfactorily the X-ray spectrum with a thermal emission of a hot, ionized, optically thin plasma model at kT=3.9 keV. Fig. \ref{fig:hd14-spectrum} shows the data and fitted spectrum of HD~141926 (see \S 4.2 for details).  

\subsection{HD~249179, HD~65663 and BD+53 2262  }

None of these three sources were detected during our survey. The X-ray emission, if present, was below the \sax\ detectable limit. Upper limits for the fluxes are given in Table \ref{tab:fluxes}. 

\label{subsect:others}

\section{Discussion}
\label{sect:discussion}

\subsection{HD~110432}

This B0.5IIIe star (Codina et al. 1984) has been proposed as a possible
  optical counterpart of the X-ray source 1H1249$-$637. The unabsorbed X-ray
  luminosity of this object during the pointed observation was $L_{2-10\rm
  keV}\simeq 3.4\times 10^{32}$ erg s$^{-1}$, assuming a distance of 301
  pc (ESA 1997) or $L_{2-10\rm keV}\simeq 7\times 10^{32}$ erg s$^{-1}$
  assuming a distance of 430 pc (Codina et al. 1984). This luminosity could
  be generated either by coronal emission of the Be star or accretion onto
  the compact companion. In the first case the light curve modulation would
  arise from self occultation of the hot spot by the Be star. However, this
  leads to an unacceptably large rotation velocity at the equator. Indeed
  Barrera, Mennickent and Vogt (1991) found a photometric period of 1.77
  days which can be regarded as the rotation period of the star (Balona
  1995). Assuming a stellar radius of $\sim$15 R$_{\odot}$ for a B0.5III
  star (Vacca et al. 1996) a rotational velocity of 430 km s$^{-1}$ can be
  deduced. This is in agreement with the velocity deduced by Codina et
  al. (1984), namely $v\sin i$=360 km s$^{-1}$ for an inclination angle of
  $i=58^{o}$ and consistent with the absence of sharp eclipses in the light
  curve. This period (1.77 d=15.3$\times 10^{4}$ s) is an order of
  magnitude larger than that measured from the X-ray light curve P$\sim
  0.16$ d. Furthermore, the X-ray luminosity is an order of magnitude
  larger than that expected for a B0.5III star. The temperature deduced
  from the spectral fitting ($kT\sim$ 11~keV) is higher than that measured
  from coronal emission of OB stars which is typically within the range of
  0.5-2.5~keV (Pallavicini et al. 1981). This clearly rules out the coronal
  emission as the origin of the observed X-ray emission. On the  other
  hand, the possibility that the light curve modulation comes from the partial occultation of the compact object during the orbital motion is clearly ruled out by Kepler's third law for the pulsation period and the radius of a B0.5III star. So we are left with partial occultation of the hot spot on the surface of the compact object during spin as a likely origin of the X-ray pulsation. 

The X-ray source was first detected by the {\it HEAO-1} experiment, with a flux
F$_{2-10\rm keV}\simeq 4.7\times 10^{-11}$ erg cm$^{-2}$ s$^{-1}$ (Tuohy
et al. 1981), while the unabsorbed flux detected by BeppoSAX is of the same order
F$_{2-10\rm keV}\simeq 3.2\times 10^{-11}$ erg cm$^{-2}$ s$^{-1}$. It
has not been detected by other probes due to its faintness. Indeed, it is
below the threshold limit of {\it RXTE ASM} ($\sim $20~mCrb). It was observed,
however, during a pointed observation with {\it ROSAT} and detected by the
{\it ROSAT}
All Sky Catalogue of bright sources, and identified as 1RXS J124250.1
$-$630332. Therefore, this source is probably a persistent low luminosity
source. Such an emission could be produced by a neutron star in very wide
orbit accreting matter from the stellar wind. The observed X-ray
luminosity is, however, an order of magnitude smaller than that usually observed in
well established neutron star Be/X-ray binaries during low or quiescent
states, $\sim 10^{33-34}$ erg s$^{-1}$ (Negueruela 1998). The luminosity
falls, however,  within the predicted range ($10^{29-33}$ erg s$^{-1}$) of wind accreting white dwarfs (Waters et al. 1989). 

The spectrum shows clearly an emission line at $\sim$ 6.8 keV as expected
from a highly ionized plasma. Owing to the large equivalent width of this
line ($\sim$ 600$-$700 eV), we interpret it as an unresolved blend of the
Fe XXV (6.70 keV) and Fe XXVI (6.97 keV) lines. This equivalent width is
much larger than those found in neutron star systems ($\leq$ 200 eV) but it
is of the order of magnitude of those found in some cataclysmic variables. The iron line at $\sim 6.4$ ~keV often exhibited by many neutron star binaries is not detected here while an emission feature around $\sim 8.4$ ~keV is observed.

The pulse period is much longer than those usually found for neutron stars, but lies within the typical values for cataclysmic variables. The spin period would imply a P$_{\rm orb}$ of the order of $10^{2}$ days (Apparao 1994b). This period agrees with the peak of the P$_{\rm orb}$ distribution computed by Raguzova (2000) for Be+WD systems. A binary system with a magnetized white dwarf in a wide orbit accreting from the stellar wind would produce persistent X-ray emission with luminosities of the order of $10^{29-33}$ erg s$^{-1}$ (Waters et al. 1989) in perfect agreement with our data.     

The $E(B-V)=0.40$ of the optical counterpart (Codina et al. 1984) implies an absorption of N$_{H}\sim 0.3\times 10^{22}$ cm$^{-2}$ (Ryter et al. 1975). However, the absorption deduced from the X-ray data fits is much higher $\sim (1.1-1.4)\times 10^{22}$ cm$^{-2}$.
Therefore, an important amount of circumstellar material should be present near the X-ray source. This would justify the presence of strong emission lines and is consistent with the system being a Be/X-ray binary.

\subsection{HD~141926}

The Be star HD~141926 has been proposed as a possible optical counterpart of the X-ray source 1H1555$-$552. We have observed it in the 1.8-10 keV
energy range, using \sax. The present study has allowed us to 
confirm, for the first time, its  nature as a  Be/X-ray emitting system, as
was suspected from early observations with the {\it HEAO-1} experiment.

As explained in section 3.2, the non-thermal models give unsatisfactory fits to our data. Good fits are obtained with thermal emission models although the absorption column remains undefined.

To further constrain the physical parameters of the source, we have studied
the optical data available in the literature. Garrison et al. (1977)
  quoted a B2III::npe spectral type. These authors warn that this classification
should be handled with due caution since it is based on a single
spectrum. Later, Reed and Beatty (1995) quote a B2nne spectral type. The photometric colors are $(B-V)=0.559$ and
$(U-B)=-0.465$. This allows us to compute the reddening free parameter $Q$
and use the calibration of Halbedel (1993) to deduce a spectral type of
B1. We
will assume a spectral type B1.5. Assuming mean values of intrinsic $(B-V)$
and $M_{\rm V}$ as given by Schmidt-Kaler (1982) for this spectral type and
luminosity classes III or V, an interstellar reddening of $E(B-V)=0.81$ can
be derived as well as distances of $\sim$ 610 pc for a class V star or $\sim$
1150 pc for a class III star, keeping in mind that these are rough
estimates. The interstellar reddening gives an absorption column of $\rm
N_{H}\sim 0.55\times 10^{22}$ cm$^{-2}$ (Ryter et al. 1975). Since in Be
type stars an extra amount of reddening is expected to come from
circumstellar material, this would represent a lower limit for $\rm
N_{H}$. Thus, we have constrained the absorption column in the fits by
setting this lower bound. Good fits are obtained for a hot thin plasma
under MEKAL or Raymond interpretations. The absorption column remains at
its lower bound with an upper limit at $2.66\times 10^{22}$ cm$^{-2}$,
implying that little circumstellar matter is present near the X-ray source. The parameters are given in Table 5.

\begin{table}
\caption[]{Model parameters for HD~141926}
\begin{flushleft}
\begin{tabular}{lr}
\hline\noalign{\smallskip}
Parameter & Value \\
\noalign{\smallskip\hrule\smallskip}

{\bf power law} &  \\
 & \\
 N$_{\rm H}$& undefined \\
$\alpha$ & 1.92$\pm$ 0.01   \\
$\chi^{2}_{\rm r}$(dof) & 1.45(4)\\
 & \\
{\bf MEKAL} & \\
 & \\
N$_{\rm H}(10^{22} \rm cm^{-2}$)& 0.55 \\
kT~(keV) & 3.67 $\pm$ 1.50  \\
$\chi^{2}_{\rm r}$(dof) & 0.97(4)\\
 & \\
{\bf Raymond} & \\
 & \\
N$_{\rm H}(10^{22} \rm cm^{-2})$& 0.55 \\
kT~(keV)& 4.08 $\pm$ 1.62  \\
$\chi^{2}_{\rm r}$(dof) & 0.93(4)\\

\noalign{\smallskip\hrule\smallskip}
\end{tabular}
\end{flushleft}
\label{tab:mod2}
\end{table}

We can therefore describe the X-ray spectrum of HD~141926 by a hot, diffuse
thin plasma at $kT\simeq 3.9\pm$0.1 ~keV (where the value and the error
correspond to the weighted mean) and $\rm N_{H}\sim 0.55\times 10^{22}$ cm$^{-2}$. There is no need to invoke any of the non-thermal components characteristic of emission from accretion onto neutron stars. 

The unabsorbed 2--10 ~keV flux for both models is 1.3$\times 10^{-12}$ erg
s$^{-1}$ cm$^{-2}$. For the distances estimated above, this would
translate into a luminosity of $L_{\rm X}\simeq 5.8\times 10^{31}$ erg
s$^{-1}$ for a class V star and $L_{\rm X}\simeq 2.1\times 10^{32}$ erg
s$^{-1}$ for a class III star, respectively. These X-ray fluxes are
 larger than those observed in B1-B2/V-III stars due to coronal
emission (Pallavicini et al. 1981) and would confirm this object as a low
luminosity Be/X-ray binary. Indeed, the bolometric luminosities for a B1.5
star are 4.2$\times 10^{37}$ erg s$^{-1}$ for a class V and 1.1$\times
10^{38}$ erg s$^{-1}$ for a class III (Schmidt-Kaler, 1982). The
expected X-ray coronal emission luminosities (Pallavicini et al. 1981) are
$L_{\rm X}=10^{-7}L_{\rm bol}=4.2\times 10^{30}$ erg s$^{-1}$ (class V) or $1.1\times 10^{31}$ erg s$^{-1}$ (class III), i.e., an order of magnitude smaller than those observed. Furthermore the measured temperature ($kT\sim$ 4 ~keV) is higher than that expected from coronal emission in early type B stars. These facts strongly suggest that HD~141926 is a low luminosity Be/X-ray binary.  
    
\subsection{DM+53~2262}

This Be star is a proposed optical counterpart of the X-ray source
1H1936$+$54. It has never been observed by imaging X-ray telescopes since
its detection during the {\it HEAO-1} survey. Our observations with \sax\
show negative results. It is completely absent in the MECS band (2--10
~keV) but could be at the very limit of detection in the LECS image
(0.1--2~keV). The upper limit of the 0.1--2~keV flux is 0.54$\times
10^{-12}$ erg s$^{-1}$ cm$^{-2}$.  This lack of detection does not exclude
its nature as a Be/X-ray binary because it could be in a quiescent state
usually found in many Be/X-ray systems. Indeed, its optical properties,
namely an early spectral type (O9--B0), large H$\alpha$ equivalent width,
and large photometric variations are compatible with this object being a
long period Be/X-ray binary (Martinez \& Fabregat, priv. comm.). In such
systems, the accretion of matter on to the neutron star can be halted by
centrifugal inhibition which occurs when the corotation radius equals the
Alfv\`{e}n radius of the neutron star magnetic field (Waters et
al. 1989b). This source is relatively close. At a distance of $\sim$22 pc
(ESA, 1997) the 2--10 keV X-ray luminosity should be $L_{\rm X}<9\times 10^{27}$ erg s$^{-1}$ which, as a matter of fact, is several orders of magnitude lower than the expected coronal emission for an early type star. 

It is worth noting that, in the MECS image, there is another source $\sim$
25\arcmin\ towards the SE, identified as 1RXS J193527.8$+$534546. This
source is inside the large error box of the {\it HEAO-1} survey of Tuohy et al. and could lead to a misidentification of DM+53~2262 as an X-ray source.  As the available data (optical and X-ray) for this source is very sparse, we did not carry the analysis any further in this paper.

\subsection{HD~65663 and HD~249179}

These sources have been proposed as possible optical counterparts of the X-ray sources 1H0749$-$600 and 1H0556$+$286 respectively. Both are included in the catalogue of HMXRBs of Liu et al. (2000). They have not been detected during our survey and upper fluxes to their emission are listed in Table 1. As in the previous case, these non detections do not exclude them as possible Be/X-ray binaries. However, both are late type stars. This argues against these objects being Be+neutron star binaries since all well established Be+NS systems have spectral types earlier than B3 in a spectral distribution which is completely different from that of isolated Be stars (Negueruela 1998). Thus, these two systems are almost certainly not neutron star accreting Be/X-ray binaries. Furthermore, it is very unlikely that these sources are Be+WD systems. Indeed, as argued by Negueruela (1998), the centrifugal inhibition of accretion would not work in these systems, so they should be persistent low luminosity sources. Owing to the lack of detection with \sax\ we consider it very likely that they are not X-ray binaries at all.       

\section{Conclusions}

We have analyzed data for five Be/X-ray binary candidates from the list proposed by Tuohy et al. (1988). We confirm HD~110432 (= 1H1249$-$637) and HD~141926 (= 1H1555$-$552) as low luminosity Be/X-ray binaries. In both cases, the spectra can be well fitted by thermal emission of a hot, optically thin, plasma. Such an emission is more consistent with systems harbouring white dwarfs instead of neutron stars as the compact objects. In the case of HD~110432 we have found an X-ray pulse period of $\sim 1.42\times 10^{4}$ s which, if confirmed by other observations, would make this object the best Be+WD candidate found to date. 

The other three objects present in our survey HD~65663, HD~249179 and
BD+53~2262 did not show detectable X-ray emission. These sources where
proposed as Be/X-ray binaries on the ground of positional coincidence with
emission line stars within the large error boxes of {\it HEAO-1} experiment
but no detection in X-rays has been reported since. BD+53~2262 has optical characteristics which are compatible with a long orbital period Be+NS binary. Such a system could present periods of quiescence due to the centrifugal inhibition of accretion mechanism. The other two objects in our sample are late type Be stars. They are almost certainly not Be+NS systems. Since the mechanism of centrifugal inhibition of accretion should not work in Be+WD systems, the possibility of a quiescent state is unlikely. We conclude therefore that, most probably, they are not X-ray binaries.

\begin{acknowledgements}
%always
The \sax\ satellite is a joint Italian-Dutch programme. JMT acknowledges
the research grants GR00-182 and AYA2000-1581-C02-02. This research has
made use of the SIMBAD database, CDS Strasbourg, France. We thank the
referee, M. Coe, for his constructive criticism. We give credit to
M. Maisack who defined the original list of sources for this \sax\
survey. We thank A. Parmar and the \sax\ LECS team for fruitful
discussions.

%% if you want
%We thank the staffs of the \sax\ Science Data and
%Operations Control Centers for help with these observations. 
%% if there are any RF authors
%M.~Guainazzi acknowledges an ESA Fellowship. 
%% if you use heasarc data
%This research has made use of data obtained through the High Energy 
%Astrophysics Science Archive Research Center Online Service, provided 
%by the NASA/Goddard Space Flight Center.
%% if you use ASM data maybe you want?
%We thank the RXTE instrument teams at MIT and NASA/GSFC for 
%providing the All-Sky Monitor Light Curves. 

\end{acknowledgements}

\end{document}